\documentclass[12pt,a4paper]{article}

\usepackage{mathtools}
\usepackage{amsfonts}
\usepackage{braket}
\usepackage{amsmath}

\usepackage{graphicx}
\usepackage{caption}
\usepackage{subcaption}

\numberwithin{equation}{section}

\begin{document}

\font\upright=cmu10 scaled\magstep1
\font\sans=cmss12


\thispagestyle{empty}
\rightline{DAMTP-2017-44}
\vskip 3em
\begin{center}
{{\bf \Large An Alpha Particle Model for Carbon-12
}} 
\\[15mm]

{\bf \large J.~I. Rawlinson\footnote{email: jir25@damtp.cam.ac.uk}} \\[1pt]
\vskip 1em
{\it 
Department of Applied Mathematics and Theoretical Physics,\\
University of Cambridge, \\
Wilberforce Road, Cambridge CB3 0WA, U.K.}
\vspace{12mm}

\abstract{We introduce a new model for the Carbon-12 nucleus and compute its lowest energy levels. Our model is inspired by previous work on the rigid body approximation in the \(B=12\) sector of the Skyrme model. We go beyond this approximation and treat the nucleus as a deformable body, finding several new states. A restricted set of deformations is considered, leading to a configuration space \(\mathcal{C}\) which has a graph-like structure. We use ideas from quantum graph theory in order to make sense of quantum mechanics on \(\mathcal{C}\) even though it is not a manifold. This is a new approach to Skyrmion quantisation and the method presented in this paper could be applied to a variety of other problems.}

\end{center}


\vskip 60pt
\vskip 5pt

\vfill
\newpage
\setcounter{page}{1}
\renewcommand{\thefootnote}{\arabic{footnote}}


\section{Motivation}

\begin{figure}[h!]
	\centering
	\begin{subfigure}{0.4\textwidth}
		\includegraphics[width=\textwidth]{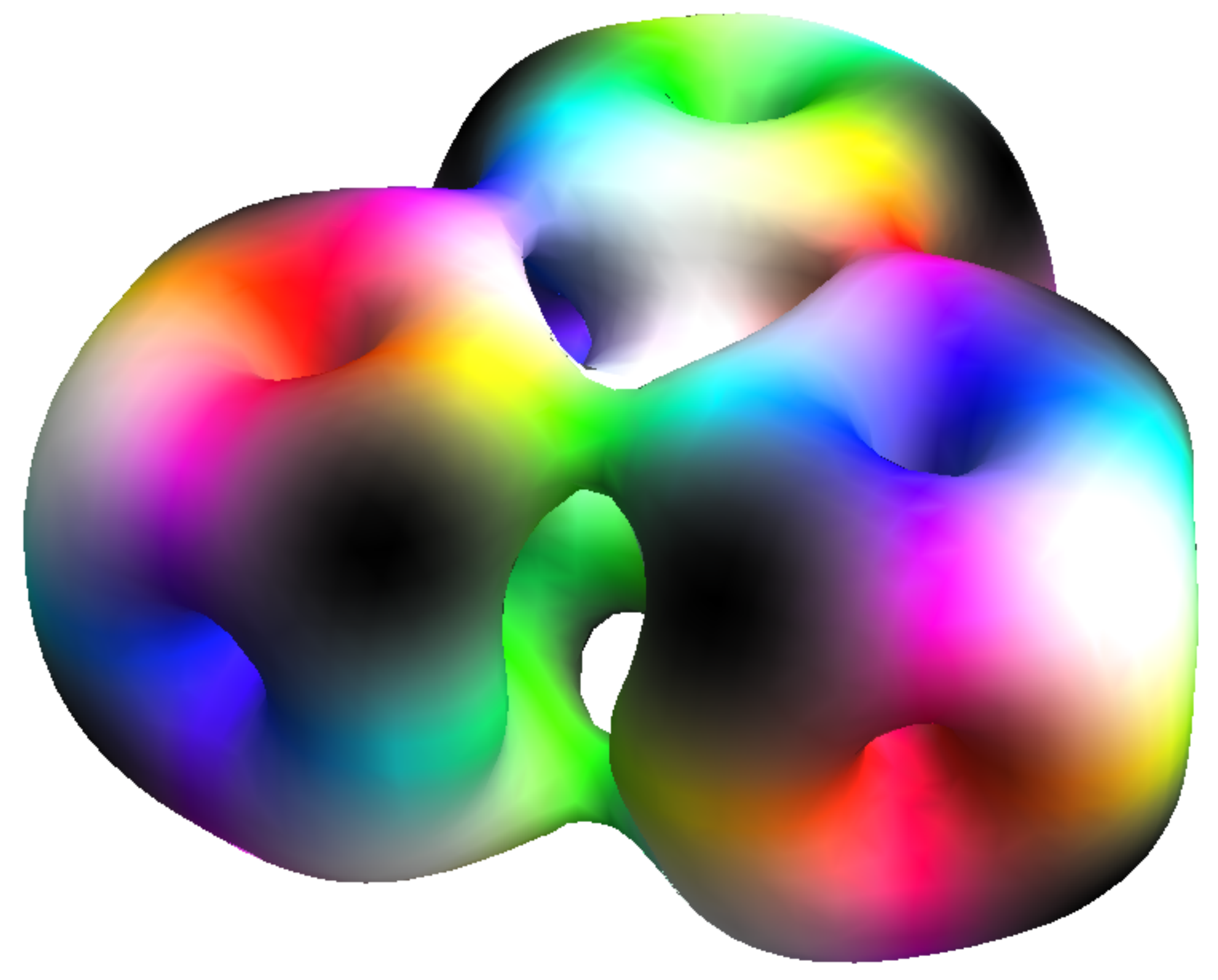} 
		
	\end{subfigure}
	\begin{subfigure}{0.55\textwidth}
		\includegraphics[width=\textwidth]{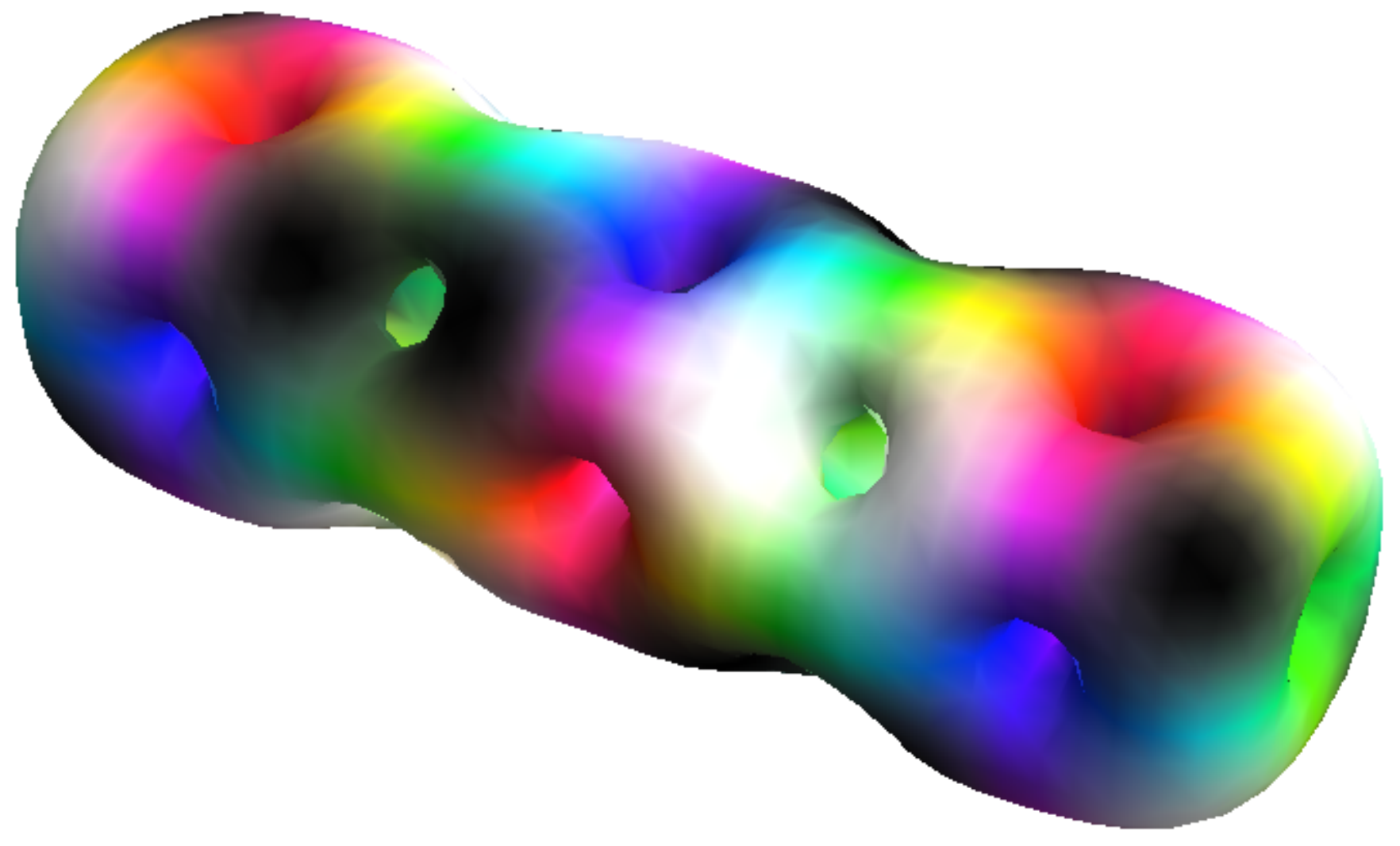} 
		
	\end{subfigure}

	\caption{\(B=12\) Skyrmions with \(D_{3h}\) symmetry (left) and \(D_{4h}\) symmetry (right). Figures courtesy of Dankrad Feist.}
	\label{fig1}
\end{figure}

The Skyrme model has two well-known Skyrmion solutions with Baryon number \(B=12\). They have symmetry groups \(D_{3h}\) and \(D_{4h}\) and can be viewed as three \(B=4\) Skyrmions (analogous to alpha particles) arranged in an equilateral triangle and in a linear chain respectively (Figure \ref{fig1}). When quantized individually as rigid bodies, each contributes a rotational band to the energy spectrum. The allowed spin and parity combinations of the quantum states appearing in each band are determined by the corresponding symmetry group (\(D_{3h}\) or \(D_{4h}\)).

The energy levels found by this approach \cite{mantonlau} match some of the experimentally observed Carbon-12 states \cite{TUNL}. The \(D_{3h}\)-symmetric Skyrmion gives rise to a rotational band with spin and parity combinations \(0^+\), \(2^+\), \(3^-\), \(4^-\), \(4^+\) (and higher spins). This is a feature of any model based on an equilateral triangle of alpha particles. A particular success of the Skyrme model prediction is the characteristic pattern \(0^+\), \(2^+\), \(4^+\) of the Hoyle band \cite{fynbo}, which arises from rotational excitations of the \(D_{4h}\)-symmetric linear chain. An alternative possibility is to identify the Hoyle band with rotational excitations of a breathing (symmetrically vibrating) equilateral triangle, as has been done within the context of the algebraic cluster model (ACM) \cite{bijkeriacello}. However, this would imply the existence of additional \(3^-\) and \(4^-\) states within the rotational band which have not been observed. It is the \(D_{4h}\) symmetry group of the Skyrmion which excludes such states. This suggests that a linear chain configuration of alpha particles could play an important role in the structure of the low-lying states of Carbon-12, as well as the more familiar equilateral triangular configuration.

However, the rigid body picture is overly restrictive: several low-lying states (seen in addition to the rigid body rotational bands) are completely missing as all of the configurations considered have a lot of symmetry. Large symmetry groups lead to severe restrictions on the possible spin and parity combinations. It is clear that we need to go beyond rigid body quantization: we need to allow the nucleus to deform. Such an approach has been successful in studying the excited states of other nuclei such as Oxygen-16 \cite{mantonhalcrowking}.

We propose a simplified model for the Carbon-12 nucleus, viewed as three point particles. Building on the ideas outlined above, we allow both the equilateral triangle and the linear chain. The difference is that they sit within a larger configuration space which includes shapes that interpolate between these two highly symmetric configurations. This extra degree of freedom allows more spin and parity combinations than rigid body quantization.

A key assumption is that the potential energy landscape is quite flat in the shape-deforming direction where the equilateral triangle becomes a linear chain. If it is not so flat, then for sufficiently low energies an analysis of small vibrations about the equilateral triangle (or about the linear chain) should be appropriate. This approximation leads to far more low-lying states than have been experimentally observed. We propose that this approximation is inappropriate: at the energies we are interested in, the triangle can deform significantly and a global approach is required.

\section{The Configuration Space, \(\mathcal{C}\)}

We start with the configuration space for three point particles,
\[\{(\mathbf{x_{1}},\mathbf{x_{2}},\mathbf{x_{3}})\in\mathbb{R}^\mathnormal{9}\}\]
equipped with the Euclidean metric. We can separate out the centre of mass motion as usual, restricting to \( \sum_{i=1}^{3}{ \mathbf{x_{i}}}=0\). We further restrict to configurations where the area of the triangle formed by the three particles has a particular shape-dependent value. Physically, this is reasonable if we assume that the energy associated with the triangle expanding compared to some equilibrium size is large compared to the energies we are interested in. The equilbrium size will be taken to vary in such a way that the ratio of the moments of inertia for the equilateral triangle and the linear chain (which determines the ratio of slopes of the corresponding rotational bands) reproduces the result from the Skyrme model.

\begin{figure}[h!]
\centering
  \includegraphics[scale=0.9]{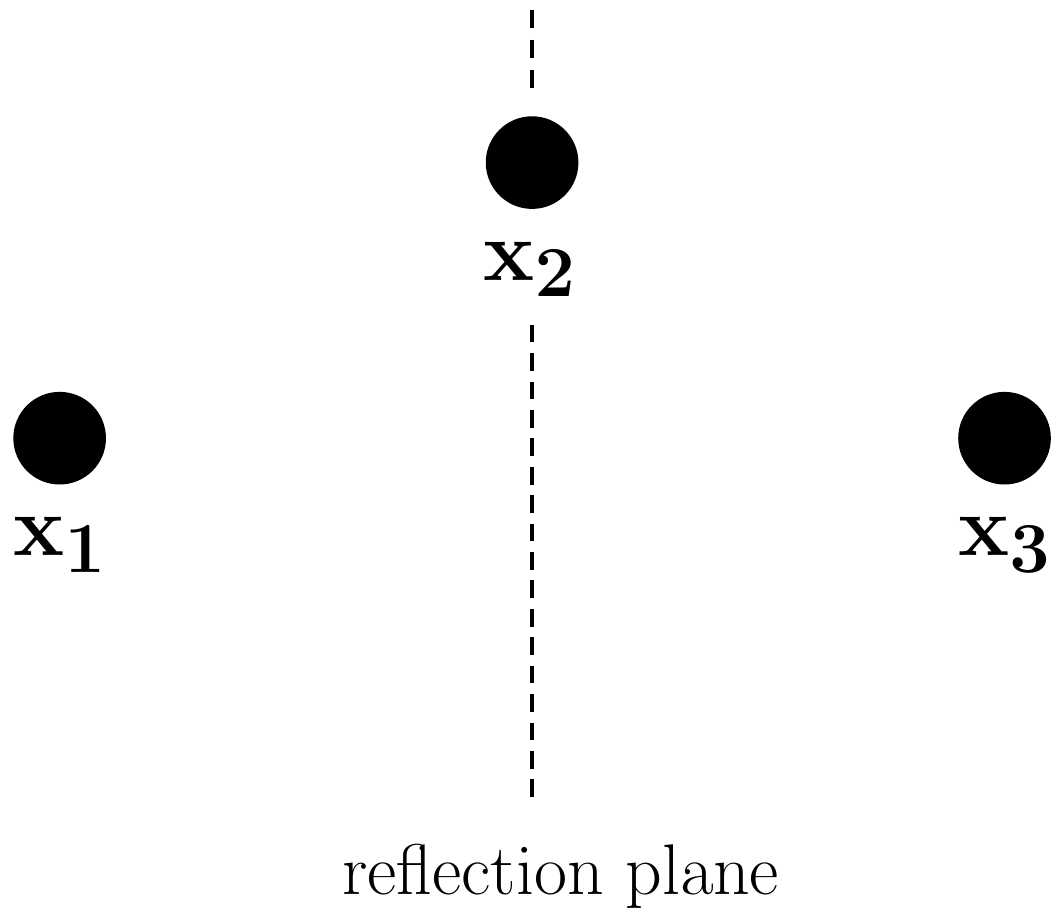}
  \caption{\(\mathcal{C}_i\) consists of isosceles configurations with particle number \(i\) lying in the plane of reflection symmetry. Thus \( \mathcal{C}_2 \) contains the above configuration. }
  \label{fig2}
\end{figure}

\begin{figure}[h!]
\centering
  \includegraphics[scale=0.9]{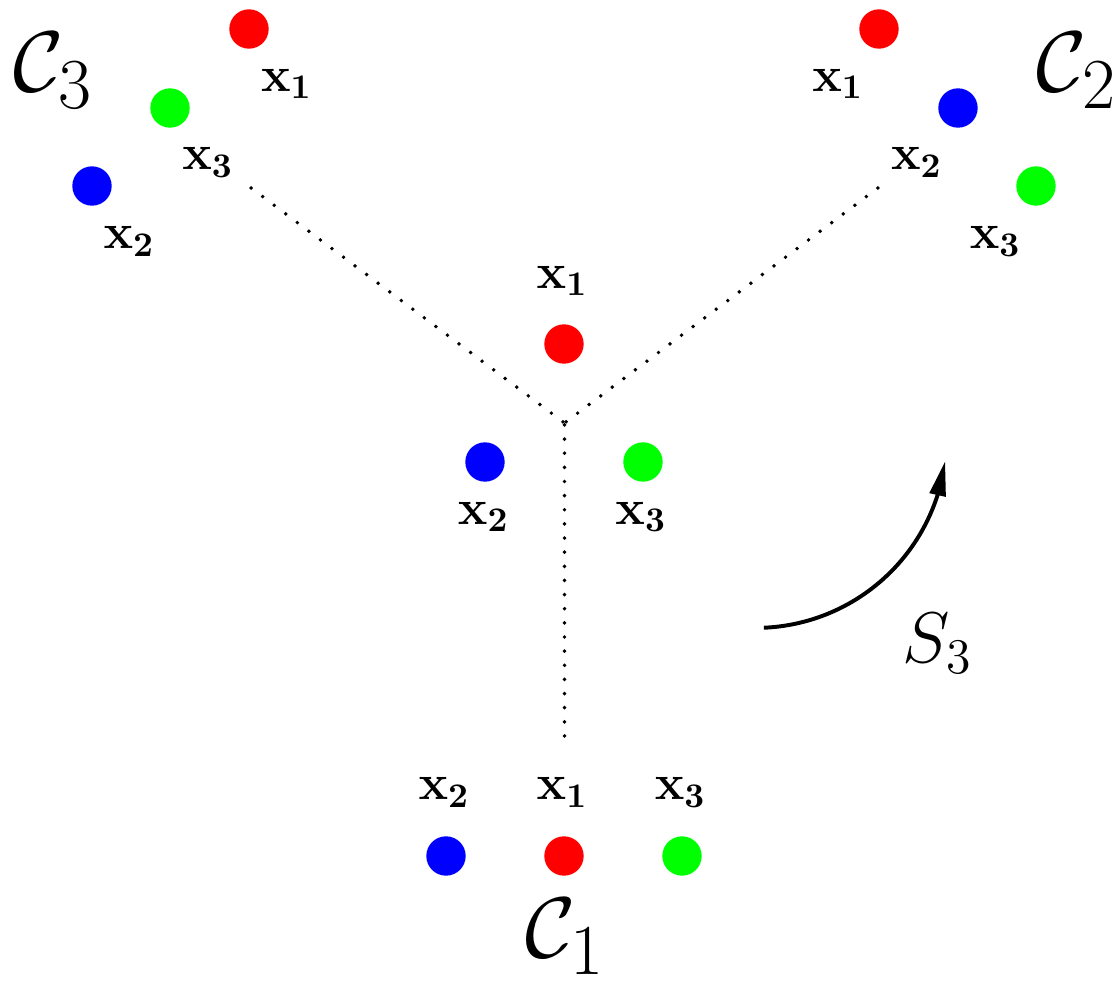}
  \caption{Graph structure of \(\mathcal{C}=\cup_{i}\mathcal{C}_i\).}
  \label{fig3}
\end{figure}

Finally, we make one further restriction: we will consider only a certain subset of configurations, in which the three particles lie at the vertices of an isosceles triangle. The shapes we restrict to are precisely those isosceles triangles which interpolate between an equilateral triangle and a linear chain. Thus we only include the direction in which the equilateral triangle becomes an obtuse triangle, assuming that changing shape in the other direction corresponds to a sharp increase in potential energy due to trying to bring two alpha particles very close together. The resulting configuration space, denoted \(\mathcal{C}\), is best pictured as the union of three 4-manifolds, \(\mathcal{C}=\cup_{i}\mathcal{C}_i\). Here \(\mathcal{C}_i\) corresponds to those isosceles configurations with particle number \(i\) lying in the plane of reflection symmetry. This is illustrated in Figures \ref{fig2} and \ref{fig3}. The \(\mathcal{C}_i\) intersect at a 3-manifold corresponding to the equilateral triangles (this intersection can be thought of as a copy of the group manifold \(SO(3)\), since this group acts on the set of equilateral triangular configurations).

The particles should be indistinguishable: this will be imposed at the quantum level by demanding that states are taken to lie in the trivial representation of the group \(S_3\) which acts on \(\mathcal{C}\) by permuting the three particles.

The structure of \(\mathcal{C}\) is reminiscent of configuration spaces appearing in quantum graph theory. We will therefore use ideas from that area to motivate our definition of quantum mechanics on \(\mathcal{C}\), thinking of the \(\mathcal{C}_i\) as edges which intersect at a vertex \cite{quantumgraphs}. Approximating a quantum system as a graph is an idea that dates back to Pauling \cite{pauling}, who studied the dynamics of free electrons in hydrocarbons by considering motion in a network with edges corresponding to carbon-carbon bonds and vertices corresponding to carbon atoms.

\section{The Metric on \(\mathcal{C}_1\)}

We begin by focusing on \(\mathcal{C}_1\). This is the only part of \(\mathcal{C}\) that we need to think about since all points in \(\mathcal{C}\) are generated by the action of \(S_3\) on \(\mathcal{C}_1\). The group of rotations in physical space acts on \(\mathcal{C}_1\) as a symmetry, so we pick coordinates \((s,\theta,\phi,\psi)\) consisting of a shape parameter \(s\) together with Euler angles \(\theta,\phi,\psi\) specifying the orientation of the shape. Corresponding to \(\theta=\phi=\psi=0\), we need to pick a set of reference orientations for each \(SO(3)\) orbit. These reference configurations give a submanifold \(\mathcal{C}_{\text{shapes}}\) which we will often refer to as shape space. We define \(\mathcal{C}_{\text{shapes}}\) as follows: let the choice \((s,0,0,0)\) correspond to the configuration
\begin{equation}\mathbf{x_{1}}=(0,s,0), \  \mathbf{ x_{2}}=\left(-\frac{1}{2}\sqrt{2-3s^2},-\frac{1}{2}s,0\right), \ \mathbf{ x_{3}}=\left(\frac{1}{2}\sqrt{2-3s^2},-\frac{1}{2}s,0\right).\end{equation}
i.e. we rotate the triangle so that particle 1 lies on the positive \(y\)-axis with the plane of reflection symmetry being the \(y-z\)-plane. The range we consider is \(s \in [0,s_{\text{max}}]\) where \(s_{\text{max}}=\frac{1}{\sqrt{3}}\). Note that \(s=0\) gives a linear chain and as we increase \(s\) we approach an equilateral triangle at \(s=s_{\text{max}}\). This particular choice of reference orientations has the nice property that the metric on \(\mathcal{C}_1\), in these coordinates, takes the simple block-diagonal form
\begin{equation}g_{\mathcal{C}_1}=\begin{pmatrix}
  \tilde{g} & 0 \\
  0 & g_{\text{rot}}
 \end{pmatrix}.\end{equation}  The reason for this simplification can be seen as follows: as we move through the reference configurations, there is enough symmetry (reflection in the \(xy\)-plane and in the \(yz\)-plane) that no angular momentum is generated by the corresponding motion of the particles. Therefore the motion in shape space decouples from the rotational motion (no cross-terms in the metric) and its only effect is through the shape-dependent moments of inertia which appear in \(g_{\text{rot}}\). The function \(\tilde{g}\) corresponds to the pull-back of the Euclidean metric on \(\mathbb{R}^\mathnormal{9}\) arising from the inclusion \(\mathcal{C}_{\text{shapes}}\xhookrightarrow{}\mathbb{R}^\mathnormal{9}\). 
 
 Once we have this metric, we can compute the associated Laplace-Beltrami operator, which is expressed in local coordinates as
 
 \begin{equation}\nabla^{2}f = \frac{1}{\sqrt{|g_{\mathcal{C}_1}|}} \partial_{i} \Big( \sqrt{|g_{\mathcal{C}_1}|}g_{\mathcal{C}_1}^{ij}\partial_{j}f \Big) .\end{equation}
 
 \section{Computing the Laplace-Beltrami Operator}
 
The formula for the Laplace-Beltrami operator involves the determinant \(|g_{\mathcal{C}_1}| = \tilde{g} |g_{\text{rot}}|\). Computing \(|g_{\text{rot}}|\) takes a little work. Let \(\mathcal{I}_{ij}\) denote the components of the moment of inertia tensor relative to the body frame of the configuration (the body frame being fixed by the standard choice of orientation as explained in the previous section). Notice that \(\mathcal{I}\) is diagonal with respect to this frame, and so the matrix form for \(g_{\text{rot}}\) with respect to coordinates (\(\theta,\phi,\psi\)) takes the simplified form  
 \begin{equation} \begin{pmatrix}
\mathcal{I}_{11} \cos^2\psi + \mathcal{I}_{22} \sin^2\psi & (\mathcal{I}_{11}-\mathcal{I}_{22}) \sin\theta \cos\psi \sin\psi & 0 \\
(\mathcal{I}_{11}-\mathcal{I}_{22}) \sin\theta \cos\psi \sin\psi & (\mathcal{I}_{11} \sin^2\psi + \mathcal{I}_{22} \cos^2\psi) \sin^2\theta + \mathcal{I}_{33} \cos^2\theta & \mathcal{I}_{33} \cos\theta \\
0 & \mathcal{I}_{33} \cos\theta & \mathcal{I}_{33} 
 \end{pmatrix}.\end{equation} All the configurations we are considering are planar, which implies that \(\mathcal{I}_{33}=\mathcal{I}_{11}+\mathcal{I}_{22}\). This simplifies the computation of the determinant of the metric and, after applying some basic trigonometric identities, we arrive at
 \begin{equation} |g_{\text{rot}}| = \mathcal{I}_{11} \mathcal{I}_{22} \mathcal{I}_{33} \sin^2\theta = |\mathcal{I}| \sin^2\theta .\end{equation} Putting everything together,
  \begin{equation}
 \begin{split}
\nabla^{2}f & = \frac{1}{\sqrt{\tilde{g} |g_{\text{rot}}|}} \partial_{i} \Big( \sqrt{\tilde{g} |g_{\text{rot}}|}g_{\mathcal{C}_1}^{ij}\partial_{j}f \Big) \\
& = \frac{1}{\sqrt{\tilde{g} |g_{\text{rot}}|}} \left( \partial_{s} \Big( \sqrt{\tilde{g} |g_{\text{rot}}|}\tilde{g}^{-1} \partial_{s}f \Big) +  \partial_{\alpha} \Big( \sqrt{\tilde{g} |g_{\text{rot}}|}g_{\text{rot}}^{\alpha\beta}\partial_{\beta}f \Big) \right) \\
& = \frac{1}{\sqrt{\tilde{g} |\mathcal{I}|}}  \partial_{s} \Big( \sqrt{\tilde{g} |\mathcal{I}|}\tilde{g}^{-1} \partial_{s}f \Big) +  \frac{1}{\sqrt{|g_{\text{rot}}|}} \partial_{\alpha} \Big( \sqrt{|g_{\text{rot}}|}g_{\text{rot}}^{\alpha\beta}\partial_{\beta}f \Big) \\
& = \widetilde{\nabla^{2}}f + \Big( \partial_{s} \log \sqrt{ |\mathcal{I}|} \Big) \tilde{g}^{-1} \partial_{s}f + \nabla^{2}_{\text{rot}}f 
\end{split}
 \end{equation} where \( \widetilde{\nabla^{2}}\) is defined by

\begin{equation} \widetilde{\nabla^{2}}f = \frac{1}{\sqrt{|\tilde{g}|}} \partial_{s} \Big( \sqrt{|\tilde{g}|}\tilde{g}^{-1} \partial_{s}f \Big) \end{equation}
and \(\nabla^{2}_{\text{rot}}\) is defined by

\begin{equation} \nabla^{2}_{\text{rot}}f = \frac{1}{\sqrt{|g_{\text{rot}}|}} \partial_{\alpha} \Big( \sqrt{|g_{\text{rot}}|}g_{\text{rot}}^{\alpha\beta}\partial_{\beta}f \Big)  . \end{equation} Roughly speaking, the first and third contributions \(\widetilde{\nabla^{2}}\) and \(\nabla^{2}_{\text{rot}}\) can be understood as kinetic energy operators associated to changes in shape and orientation respectively. They are the Laplace-Beltrami operators associated to submanifolds comprising configurations of fixed orientation and of fixed shape. The second contribution is the manifestation of the \(s\) dependence of the moment of inertia tensor \(\mathcal{I}\). It is conventional to express the operator \(\nabla^{2}_{\text{rot}}\) in terms of the vector \(\mathbf{L}\) of body-fixed angular momentum operators:
\begin{equation}
\nabla^{2}_{\text{rot}}f =  - \mathbf{L}^{\text{T}} \mathcal{I}^{-1} \mathbf{L}f.
\end{equation}
 
 \section{Quantum Mechanics on \(\mathcal{C}\)}
 \subsection{Schr\"odinger equation and symmetries}
 We are now in a position to write down the form of the Schr\"odinger equation, at least on \(\mathcal{C}_1\):
 \begin{equation} \mathcal{H} \Psi \equiv -\frac{1}{2}\nabla^{2}\Psi + V\Psi\ =E\Psi.\end{equation} As described in the previous section, it is most convenient to express the Laplace-Beltrami operator with respect to coordinates \((s,\theta,\phi,\psi)\). In these coordinates, the Schr\"odinger equation is

\begin{equation} \left( \frac{1}{2} \mathbf{L}^{\text{T}} \mathcal{I}^{-1} \mathbf{L} - \frac{1}{2} \widetilde{\nabla^{2}} - \frac{1}{2} \Big( \partial_{s} \log \sqrt{ |\mathcal{I}|} \Big) \tilde{g}^{-1} \partial_{s} \right) \Psi = (E-V)\Psi. \end{equation}  We can use rotational symmetry to simplify the problem considerably. Since (space-fixed) angular momentum \(\mathbf{J}\) is conserved, we can expand \(\Psi\) in a basis of states of fixed \(J\) and \(j_3\):

\begin{equation} \label{genexpansion}
 \Psi(s,\theta,\phi,\psi)=\sum_{l_3=-J}^{l_3=J}\chi_{l_3}(s)\ket{J,j_3,l_3}
 \end{equation}
 where \(l_3, j_3\) denote the projection of angular momentum onto the body-fixed and space-fixed 3-axes, i.e. the eigenvalues of the operators \({L}_3, {J}_3\). So states with fixed  \(J\) and \(j_3\) lie in a subspace spanned by the set \(\{\ket{J,j_3,-J},...,\ket{J,j_3,J}\}\). We will denote this subspace \(V_J\), from here on suppressing the label \(j_3\).
 
 In fact we can go further since there are additional discrete symmetries present: the system is symmetric under permutations of the three particles and under the action of parity. Note that parity is realised by a rotation through \(\pi\) about the body-fixed 3-axis since the configurations are planar. The parity transformation commutes with the permutations and so together these transformations generate a symmetry group isomorphic to \(S_3 \times C_2\) which acts on \(V_J\). Thus we can further classify states into irreducible representations of the discrete group \(S_3 \times C_2\). These are precisely the tensor products \(\rho_{S_3} \otimes \rho_{C_2}\) of irreducible representations \(\rho_{S_3}\) and \(\rho_{C_2}\) of \(S_3\) and \(C_2\) respectively. We want the particles to be indistinguishable, so we choose those representations for which the first factor is the trivial representation:  \(\rho_{S_3}=1\). If we let \(\rho_{\text{trivial}}\) and \(\rho_{\text{sign}}\) denote the (familiar) irreducible representations of \(C_2\), then this means that states within \(V_J\) can be taken to lie in representations isomorphic to \(1 \otimes \rho_{\text{trivial}}\) or \(1 \otimes \rho_{\text{sign}}\). Those transforming in the representation \(1 \otimes \rho_{\text{trivial}}\) will be referred to as positive parity states, \(J^+\), and those transforming in \(1 \otimes \rho_{\text{sign}}\) as negative parity states \(J^-\).
 
 Substituting expression (\ref{genexpansion}) into the Schr\"odinger equation leads to a system of coupled equations for the effective one-dimensional problem on shape space. The simplest case to consider is \(J=0\). Note that in this subspace only positive parity states \(0^+\) exist. \(\mathbf{L}\) acts on these states as the zero operator which leaves us with a single ordinary differential equation in \(s\):
 
 \begin{equation} \left(  - \frac{1}{2} \widetilde{\nabla^{2}} - \frac{1}{2} \Big( \partial_{s} \log \sqrt{ |\mathcal{I}|} \Big) \tilde{g}^{-1} \partial_{s} \right) \chi_{0} = (E-V) \chi_{0} .\end{equation}  Solving this equation for \(\chi_{0}\) gives the total wavefunction
 
  \begin{equation}\Psi(s,\theta,\psi,\psi)=\chi_{0}(s)\ket{0,0}.\end{equation}  For higher \(J\) the term of the form \( (\mathbf{L}^{\text{T}} \mathcal{I}^{-1}(s) \mathbf{L}) \Psi \) appearing in the Schr\"odinger equation will mix states with different \(l_3\) values, leading to systems of coupled ordinary differential equations. Let us consider \(J=2\) as an example. We begin by writing down the most general element of \(V_{2}\):
   \begin{equation}\Psi = \chi_{-2}\ket{2,-2}+\chi_{-1}\ket{2,-1}+\chi_{0}\ket{2,0}+\chi_{1}\ket{2,1}+\chi_{2}\ket{2,2}.\end{equation} Suppose we are interested in positive parity states \(2^+\). Such states lie in \(V_2\) and transform in the representation \(1 \otimes \rho_{\text{trivial}}\) of \(S_3 \times C_2\). In particular, they should transform trivially under the action of both \(\exp(i\pi L_2)\) (swapping particles 2 and 3) and \(\exp(i\pi L_3)\) (parity). So they lie in a 2 dimensional subspace spanned by the set \(\{\ket{2,0},\ket{2,-2}+\ket{2,2}\}\). In other words, the \(2^+\) states have the restricted form
     \begin{equation}\Psi = \chi_{0}\ket{2,0}+\chi_{2}(\ket{2,-2} + \ket{2,2}).\end{equation}
     It is convenient to write \(\Psi = \begin{pmatrix}
          \chi_0\\
     \chi_2
     \end{pmatrix} \) as a column vector with respect to the above basis. It can be checked using the explicit matrix representation of \( \mathbf{L}^{\text{T}} \mathcal{I}^{-1} \mathbf{L}\) with respect to this basis that  \begin{equation}(\mathbf{L}^{\text{T}} \mathcal{I}^{-1} \mathbf{L}) \Psi =  \begin{pmatrix} 
     3(\mathcal{I}^{-1}_{11}+\mathcal{I}^{-1}_{22}) & \sqrt{6}(\mathcal{I}^{-1}_{11}-\mathcal{I}^{-1}_{22}) \\ 
      \frac{\sqrt{6}}{2}(\mathcal{I}^{-1}_{11}-\mathcal{I}^{-1}_{22})  &  \mathcal{I}^{-1}_{11}+\mathcal{I}^{-1}_{22}+4\mathcal{I}^{-1}_{33} 
     \end{pmatrix} 
      \begin{pmatrix} 
     \chi_0\\
     \chi_2
     \end{pmatrix} 
    . \end{equation}The off-diagonal terms in the matrix are non-zero so we have a coupled system for the two functions \(\chi_0\) and \(\chi_2\).
    
    More generally, for spin \(J\) states the operator \(\exp(i\pi L_2)\) has the following diagonal matrix form with respect to the basis \(\{\ket{J,-J},\ldots,\ket{J,J}\}\):
    \begin{equation}\exp(i\pi L_2)=(-1)^{J}\begin{pmatrix}
     &  &  &  &  &  &  &  & (-1)^{J}\\
    &  &  &  &  &  &  & \reflectbox{\(\ddots\)} & \\
     &  &  &  &  &  & 1 &  & \\
     &  &  &  &  & -1 &  &  & \\
     &  &  &  & 1 &  &  &  & \\
     &  &  & -1 &  &  &  &  & \\
     &  & 1 & & & & & &\\
     & \reflectbox{\(\ddots\)} & & & & & & &\\
    (-1)^{J} & & & & & & & &\\
    \end{pmatrix}.\end{equation}
    This observation makes it very easy to compute a basis for allowed states of a given spin and parity, summarised in Table \ref{table1}.
    
    \begin{table}[h!]
    \centering
    \begin{tabular}{ c | l || c | l}
    \(J^{P}\) & basis & \(J^{P}\) & basis \\ \hline
    \(0^+\) & \(\ket{0,0}\) & \(0^-\) &  \\ \hline
     \(1^+\) & & \(1^-\) & \(\ket{1,-1} + \ket{1,1}\)   \\ \hline
        \(2^+\)  & \(\ket{2,-2} + \ket{2,2}\) & \(2^-\)  & \(\ket{2,-1} - \ket{2,1}\)  \\
              & \(\ket{2,0} \) &    &  \\ \hline
        \(3^+\) & \(\ket{3,-2} - \ket{3,2}\) & \(3^-\) & \(\ket{3,-3} + \ket{3,3}\) \\
        &  & & \(\ket{3,-1} + \ket{3,1}\) \\ \hline
         \(4^+\) & \(\ket{4,-4} + \ket{4,4}\) & \(4^-\) & \(\ket{4,-3} - \ket{4,3}\) \\
        & \(\ket{4,-2} + \ket{4,2}\) & & \(\ket{4,-1} - \ket{4,1}\)  \\
      & \(\ket{4,0}\) & & \\ \hline
        \(5^+\)  & \(\ket{5,-4} - \ket{5,4}\) & \(5^-\)  & \(\ket{5,-5} + \ket{5,5}\)  \\
              &  \(\ket{5,-2} - \ket{5,2}\) &    &  \(\ket{5,-3} + \ket{5,3}\)  \\
               &  &    &  \(\ket{5,-1} + \ket{5,1}\)  \\
      
    \end{tabular}
    \caption{Bases for allowed states.}
    \label{table1}
    \end{table}

\subsection{Boundary conditions}  
We can solve these equations numerically, once we have specified suitable boundary conditions on the wavefunction. The appropriate boundary conditions to impose at the intersection of the \(\mathcal{C}_i\) are not immediately obvious: following the quantum graph literature, we demand that the wavefunction is continuous at the equilateral triangle (\(s=s_{max}=\frac{1}{\sqrt{3}} \approx 0.58\)) with outgoing derivatives along the three edges (the \(\mathcal{C}_{i}\)) summing to zero. The latter condition ensures conservation of the probability current \cite{networkmodel}. By the outgoing derivative along an edge we mean the derivative in the direction orthogonal to the fibre generated by the action of the rotation group: in the case of \(\mathcal{C}_1\), for which we have constructed explicit coordinates, this just means the partial derivative with respect to the coordinate \(s\) (the block-diagonal form of the metric \(g_{\mathcal{C}_1}\) makes it clear that this direction is orthogonal to the action of rotations).
 
\begin{figure}[h!]
\centering
  \includegraphics[scale=1]{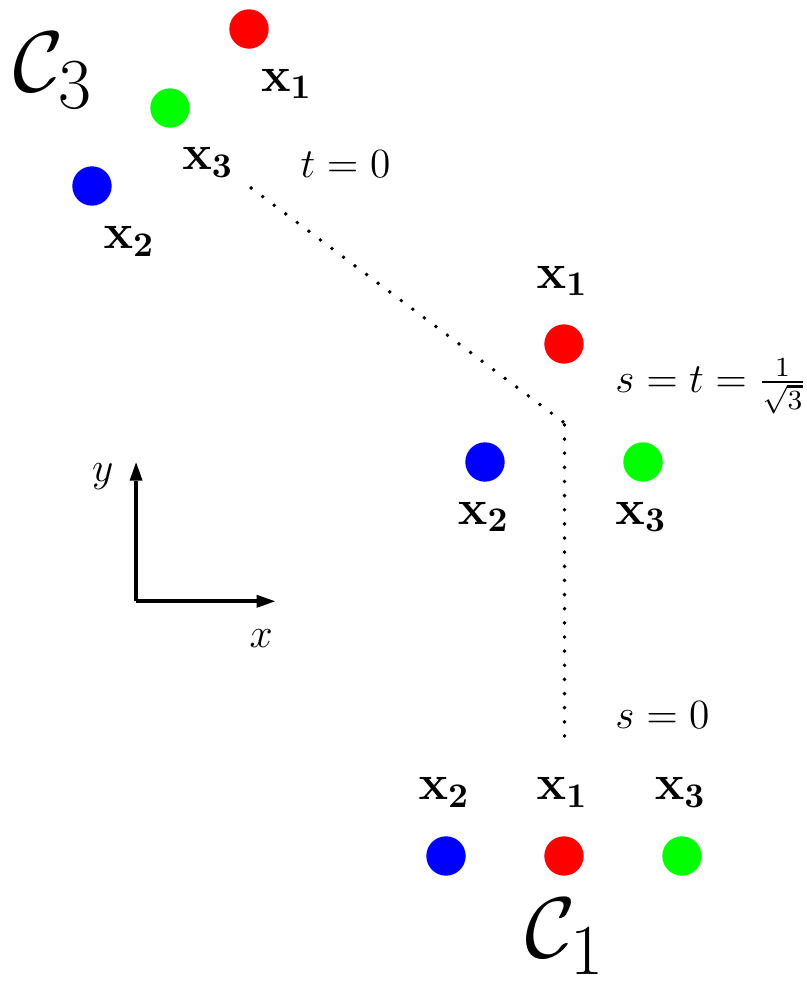}
  \caption{Cross-section of part of \(\mathcal{C}\) with \(\theta=\phi=\psi=0\).}
  \label{fig4}
\end{figure}

In our problem we have the additional requirement of indistinguishability: \(S_3\) must act trivially on our solutions. This means that the wavefunction on \(\mathcal{C}_{2}\) and \(\mathcal{C}_{3}\) is determined by the wavefunction on \(\mathcal{C}_{1}\), and so these boundary conditions are in every case equivalent to some boundary condition on the \(\chi_{l_3}\) (which are functions defined only on the smaller set \(\mathcal{C}_{1}\)). To make this clearer, consider Figure \ref{fig4}. 

Suppose we have constructed local coordinates \((t,\theta,\phi,\psi)\) on \(\mathcal{C}_3\) just as on \(\mathcal{C}_1\) and that the illustrated configurations correspond to the cross-section of \(\mathcal{C}\) given by \(\theta=\phi=\psi=0\). It is clear from the diagram that indistinguishability implies that the wavefunction on \(\mathcal{C}_3\) should be related to the wavefunction on \(\mathcal{C}_1\) by 
\begin{equation} \Psi^{\mathcal{C}_3}=\exp\left(i\frac{2\pi}{3} L_3\right) \Psi^{\mathcal{C}_1}, \end{equation}
i.e. if we rotate the configurations shown in \(\mathcal{C}_1\) by \(\frac{2\pi}{3}\) then they agree with the configurations shown in \(\mathcal{C}_3\) up to relabelling of particle number and so are physically indistinguishable. Going back to our \(2^+\) example, we had that 
\begin{equation} \Psi^{\mathcal{C}_1} = \chi_{0}\ket{2,0}+\chi_{2}(\ket{2,-2} + \ket{2,2}). \end{equation} 
So 
\begin{equation} \exp\left(i\frac{2\pi}{3} L_3\right) \Psi^{\mathcal{C}_1} = \chi_{0}\ket{2,0}+\chi_{2}(\exp\left(-i\frac{4\pi}{3}\right)\ket{2,-2} + \exp\left(i\frac{4\pi}{3}\right)\ket{2,2}). \end{equation}
Continuity of the wavefunction then implies \(\chi_{2}\) vanishes at the equilateral triangle while the derivative condition implies the derivative of \(\chi_{0}\) vanishes at the equilateral triangle. An example of such a solution can be seen later in Figure \ref{fig5}.

\section{Wavefunctions and Energy Levels}

\begin{figure}[h!]
\centering
  \includegraphics[scale=0.7]{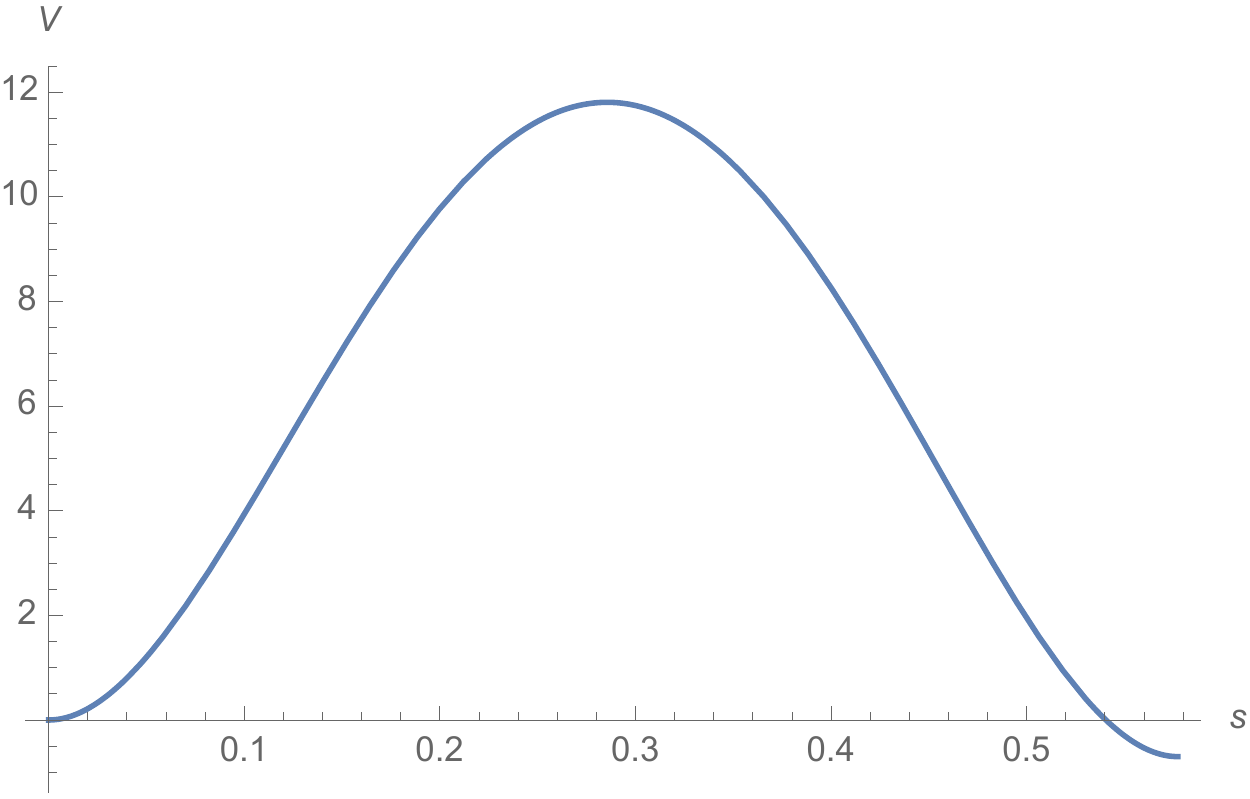}
  \caption{Potential \(V(s)\). This particular \(V\) corresponds to our final choice of parameters, and is expressed in MeV.}
  \label{fig6}
\end{figure}

We want a potential \(V(s)\) which has local minima at the linear chain and at the equilateral triangle. A simple choice is a quartic polynomial (Figure \ref{fig6}). This involves picking five coefficients, but the two conditions just mentioned together with the freedom to specify a zero point energy means that such a potential has only two free parameters. These can be thought of as the energy difference between the two minima and the height of the potential barrier between them.

We fix our energy units by matching the energy difference between the lowest-lying \(0^{+}\) and \(2^{+}\) states of Carbon-12 to experiment, and adjust the values of the two parameters in the potential to give a spectrum which is closest to the experimentally observed energies.

\subsection{Asymptotic rigid body regime}

\begin{figure}[h!]
\centering
  \includegraphics[scale=1.2]{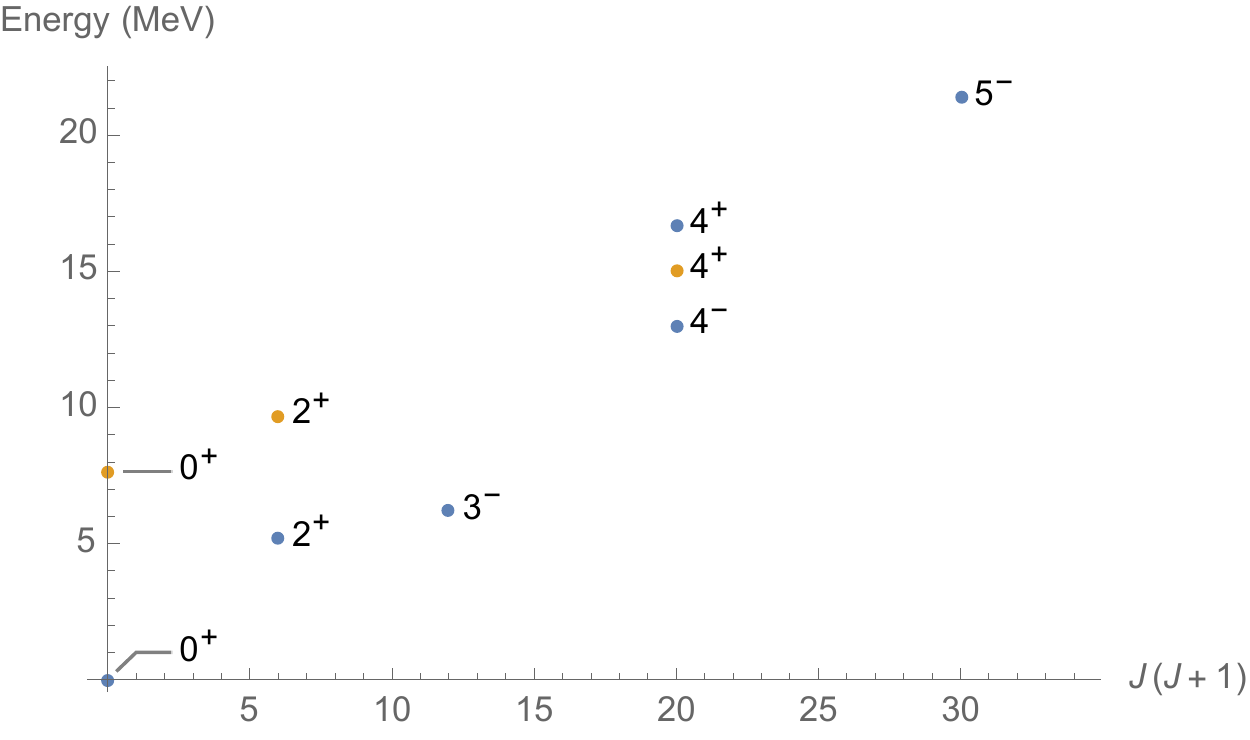}
  \caption{Energy spectrum in the rigid body regime.}
  \label{fig7}
\end{figure}

When we pick parameters such that there is a large potential barrier between the equilateral triangle and the linear chain, we recover the usual rigid body picture \cite{mantonlau}, whose spectrum is shown in Figure \ref{fig7}.

\begin{figure}[h!]
	\centering
	
	\begin{subfigure}{0.49\textwidth}
		\includegraphics[width=\textwidth]{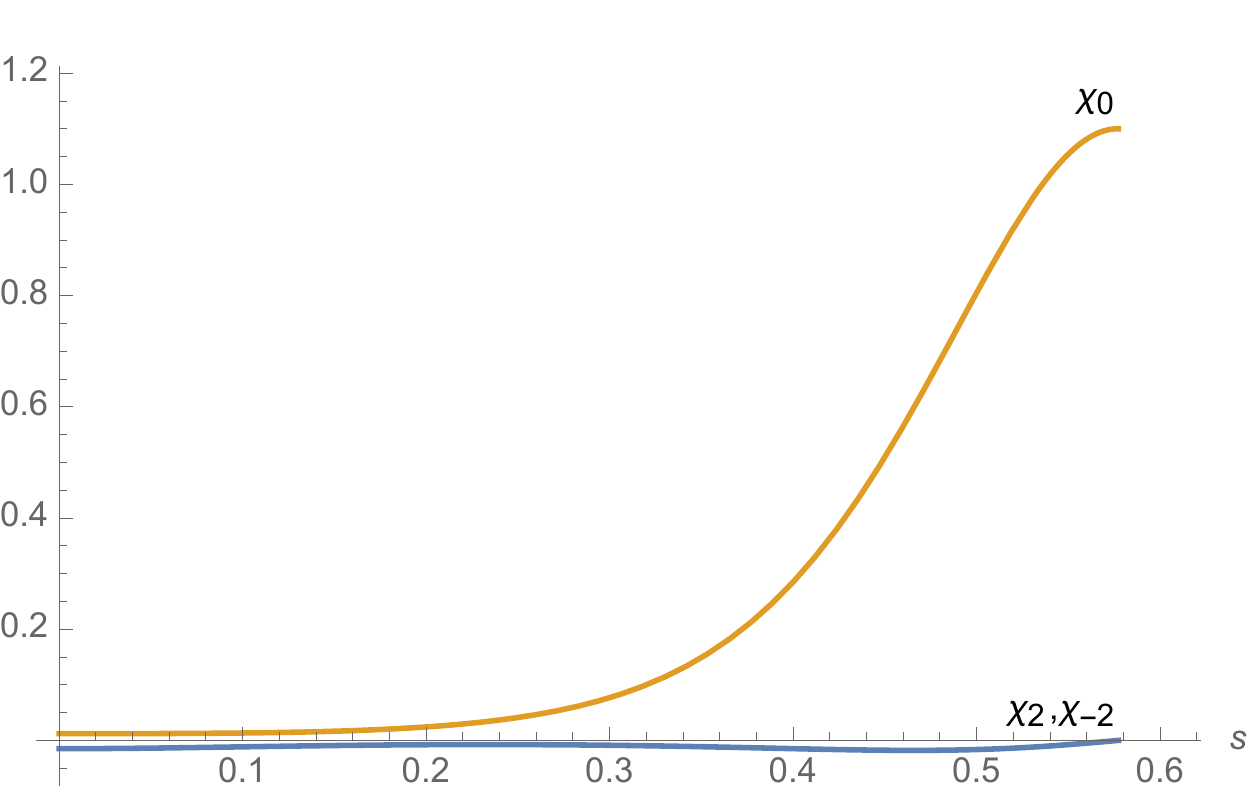} 
		
	\end{subfigure}
	\begin{subfigure}{0.49\textwidth}
		\includegraphics[width=\textwidth]{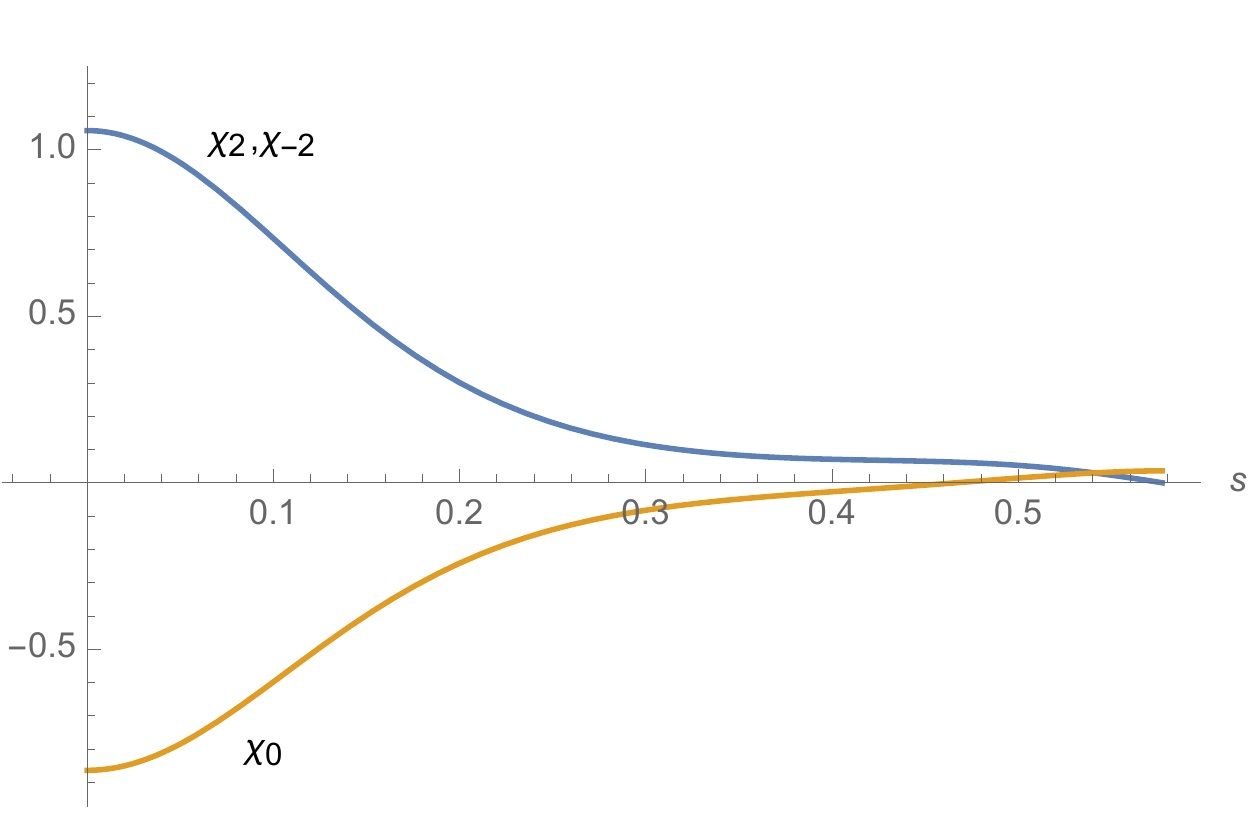} 
		
	\end{subfigure}
		
	\caption{Wavefunctions of the lowest-lying \(2^+\) states in the rigid body regime, concentrated on the equilateral triangle (left) and the linear chain (right).}
	\label{fig5}
\end{figure}

In this asymptotic regime, the wavefunctions become concentrated on either the equilateral triangle (blue, often referred to as the ground state band) or the linear chain (yellow, often referred to as the Hoyle band) for spins and parities that are allowed at those shapes. Recall that the energy levels of a symmetric top (having moments of inertia \(V_{11}=V_{22}\), \(V_{33}\)) are given by \(E=\frac{1}{2V_{11}} J(J+1) + \left(\frac{1}{2V_{33}}-\frac{1}{2V_{11}} \right) l_{3}^2\). The Hoyle band states in the plot all have body-fixed spin projection \(|l_{3}|=0\) and so the corresponding energy values lie on a straight line. The ground state band includes states with \(|l_{3}|=0\) and \(|l_{3}|=3\), with the \(|l_{3}|=0\) states lying on a straight line and the \(|l_{3}|=3\) states just below this line. Some examples of these wavefunctions are in Figure \ref{fig5}.

A shortcoming of the rigid body spectrum is the absence of the experimentally observed low-lying \(1^-\) and \(2^-\) states, both of which are known to have energies less than 15 MeV. 
There are additional states which are allowed in our model, including some new spin and parity combinations such as \(1^-\), \(2^-\) and \(3^+\). These new states have the same quantum numbers as (some of) those found by the ACM approach \cite{bijkeriacello}. However, in the rigid body regime they have high energy (outside of the range shown in Figure \ref{fig7}) since they must vanish at both the equilateral triangle and linear chain. They are peaked at intermediate configurations which have very high energy.

\subsection{Relaxing the rigid body assumption}

\begin{figure}[h!]
	\centering
	
	\begin{subfigure}{0.49\textwidth}
		\includegraphics[width=\textwidth]{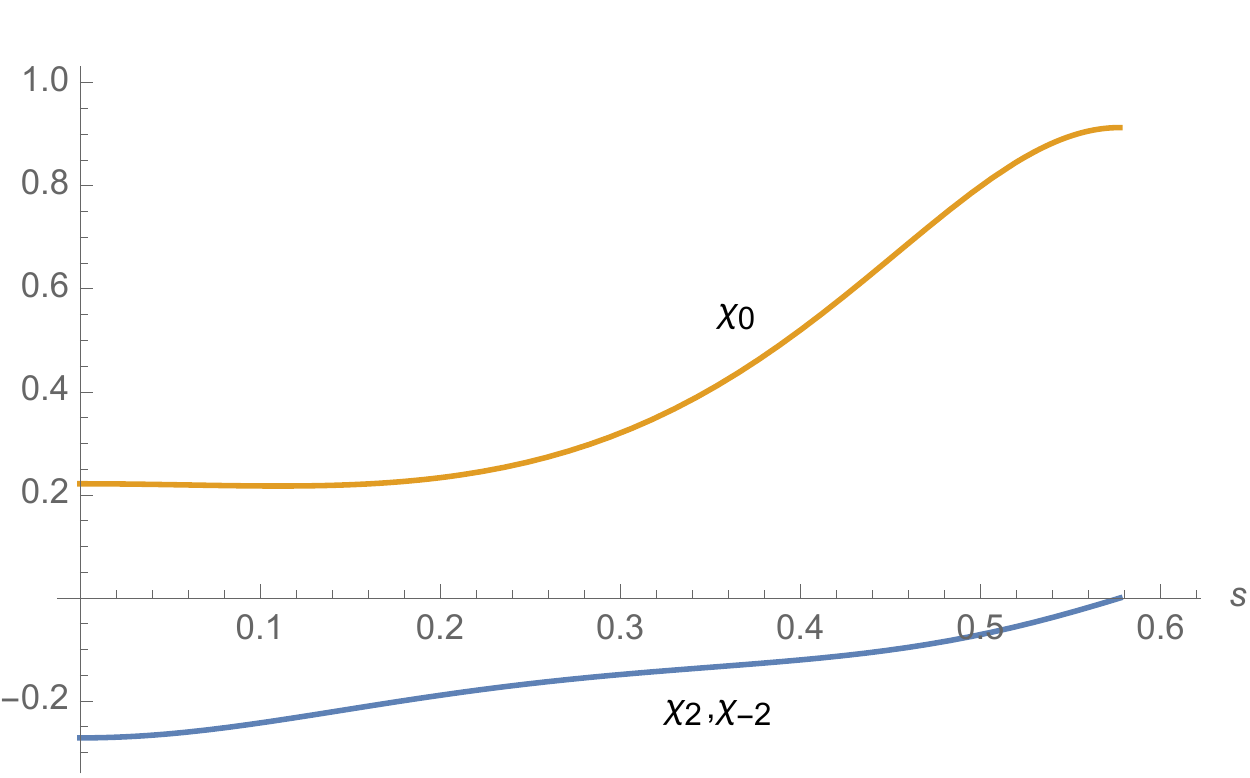} 
		
	\end{subfigure}
	\begin{subfigure}{0.49\textwidth}
		\includegraphics[width=\textwidth]{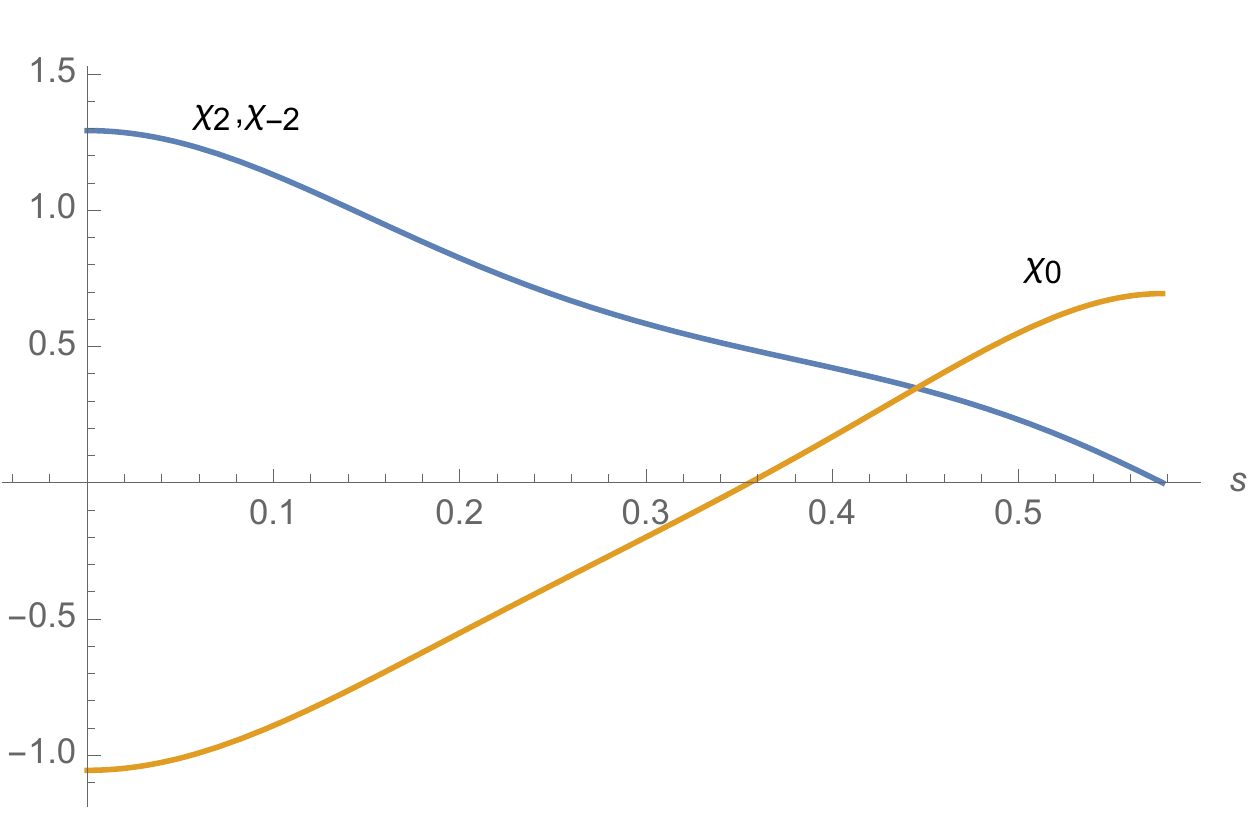} 
		
	\end{subfigure}
		
	\caption{Wavefunction of the lowest-lying \(2^+\) states: these correspond to superpositions of the equilateral triangle and linear chain states in Figure \ref{fig5}, reflecting the relaxation of the rigid body assumption.}
	\label{fig8}
\end{figure}

\begin{figure}[h!]
	\centering
	\begin{subfigure}{0.49\textwidth}
		\includegraphics[width=\textwidth]{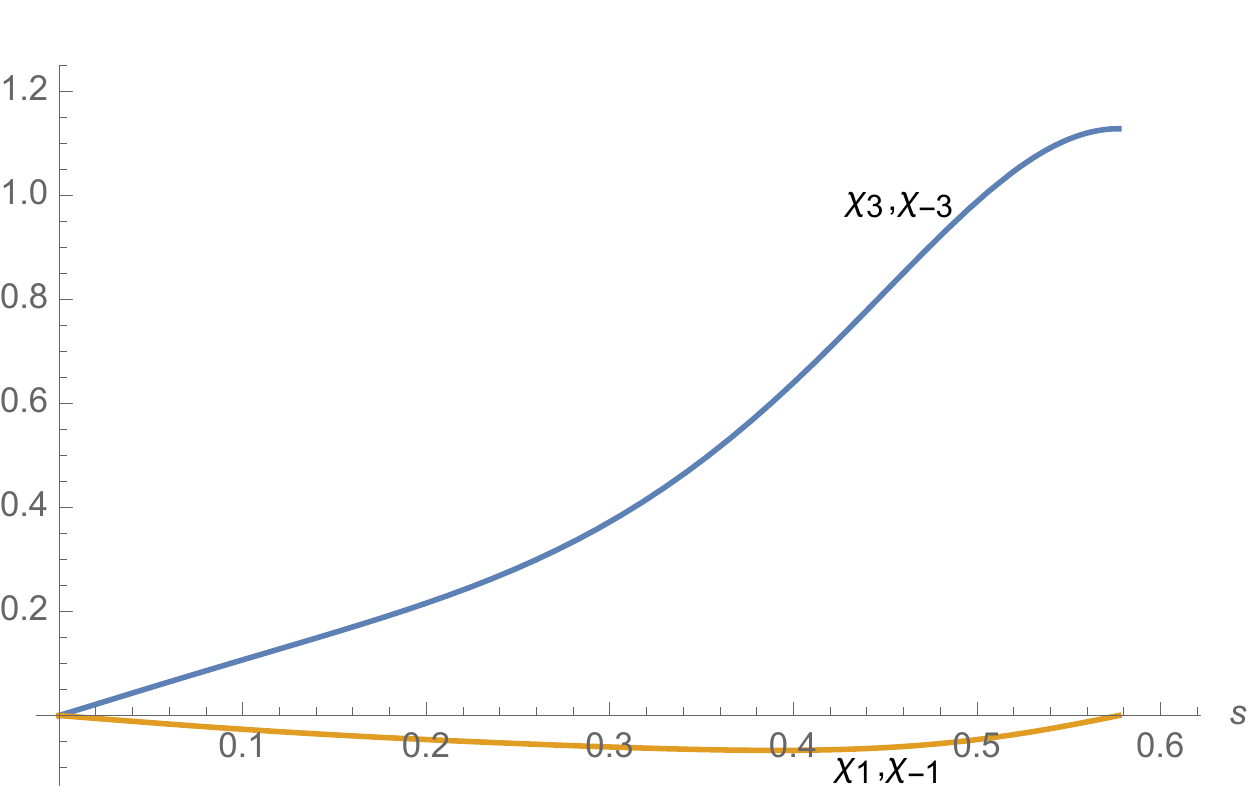} 
		
	\end{subfigure}
	\begin{subfigure}{0.49\textwidth}
		\includegraphics[width=\textwidth]{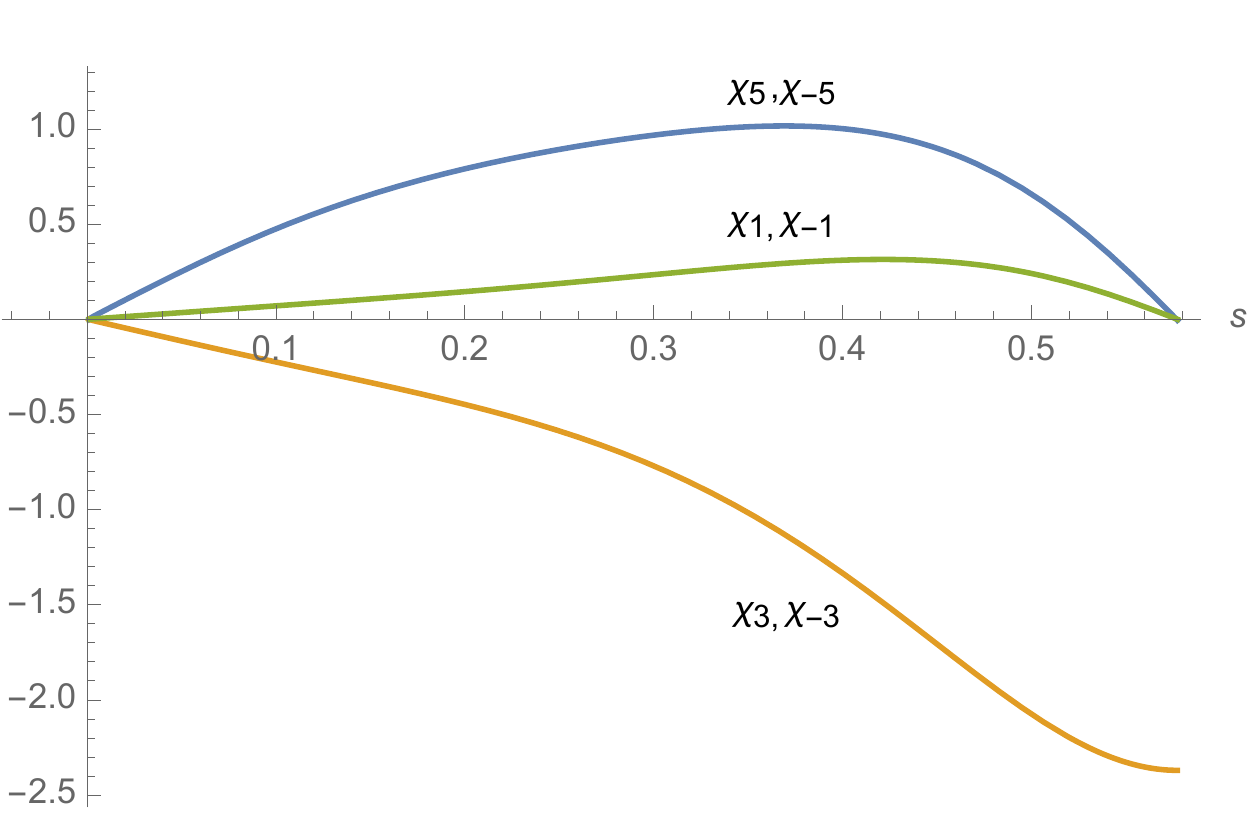} 
		
	\end{subfigure}

	\caption{Wavefunctions of the \(3^-\) state and the \(5^-\) state. Note that they must vanish at the linear chain \(s=0\).}
	\label{fig12}
\end{figure}

Starting from parameters corresponding to the rigid body regime, we consider the effect of lowering the size of the barrier. States concentrated at the equilateral triangle or the linear chain start to mix with one another if they have the same quantum numbers. This means that the \(0^+\) and \(2^+\) states concentrated at the equilateral triangle become a superposition of triangular and linear chain states (Figure \ref{fig8}). Very little mixing can occur for states such as the \(3^-\) and \(5^-\), however, as they are only allowed at the equilateral triangle and not at the linear chain (Figure \ref{fig12}).

\begin{figure}[h!]
\centering
  \includegraphics[scale=0.85]{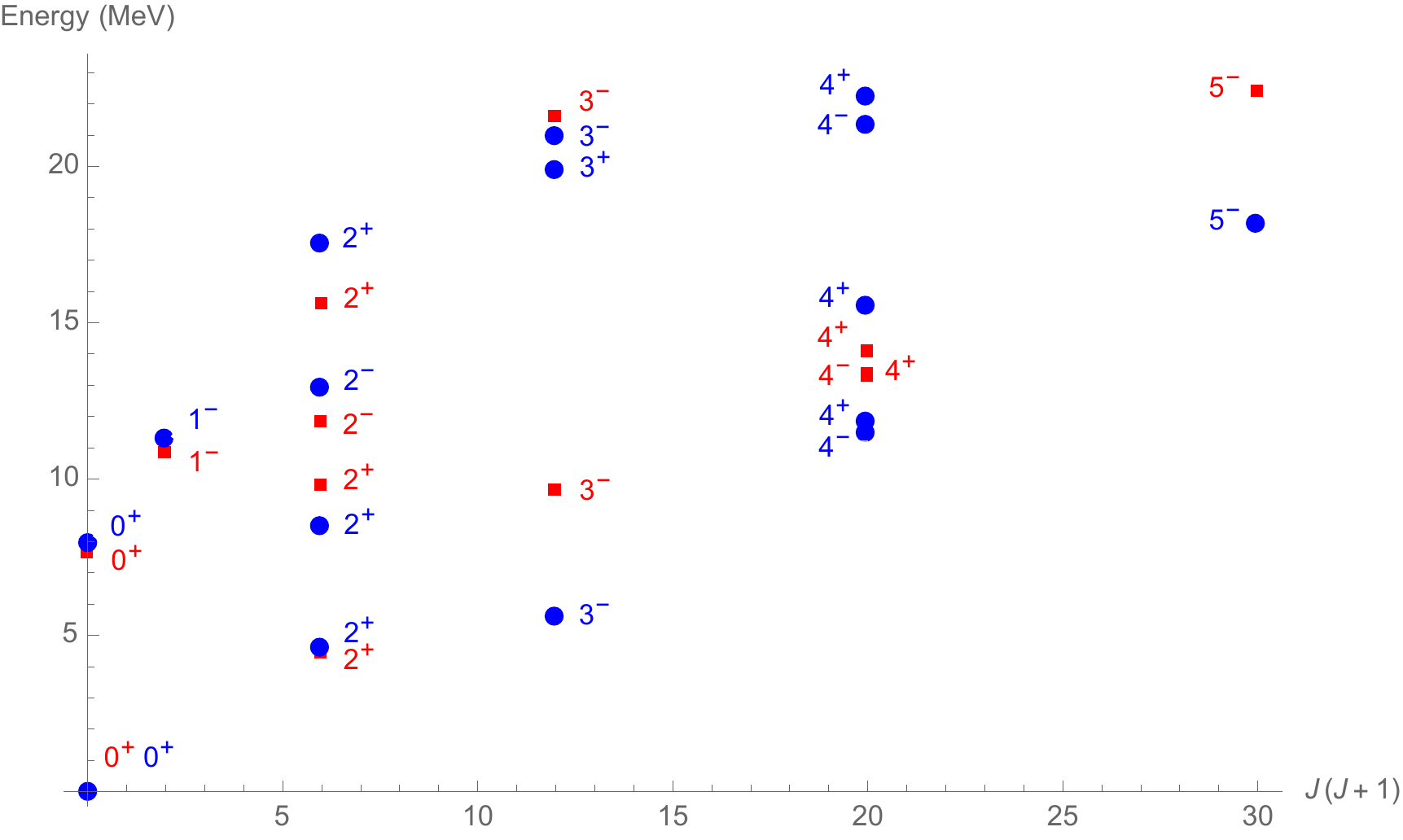}
  \caption{Spectrum of our model (blue points) compared to experimental data (red points).}
  \label{fig13}
\end{figure}

The energy of the new \(1^-\) state decreases and we stop lowering the barrier when the experimental value is reached. The final spectrum is displayed in Figure \ref{fig13}.

\begin{figure}[h!]
	\centering
	\begin{subfigure}{0.49\textwidth}
		\includegraphics[width=\textwidth]{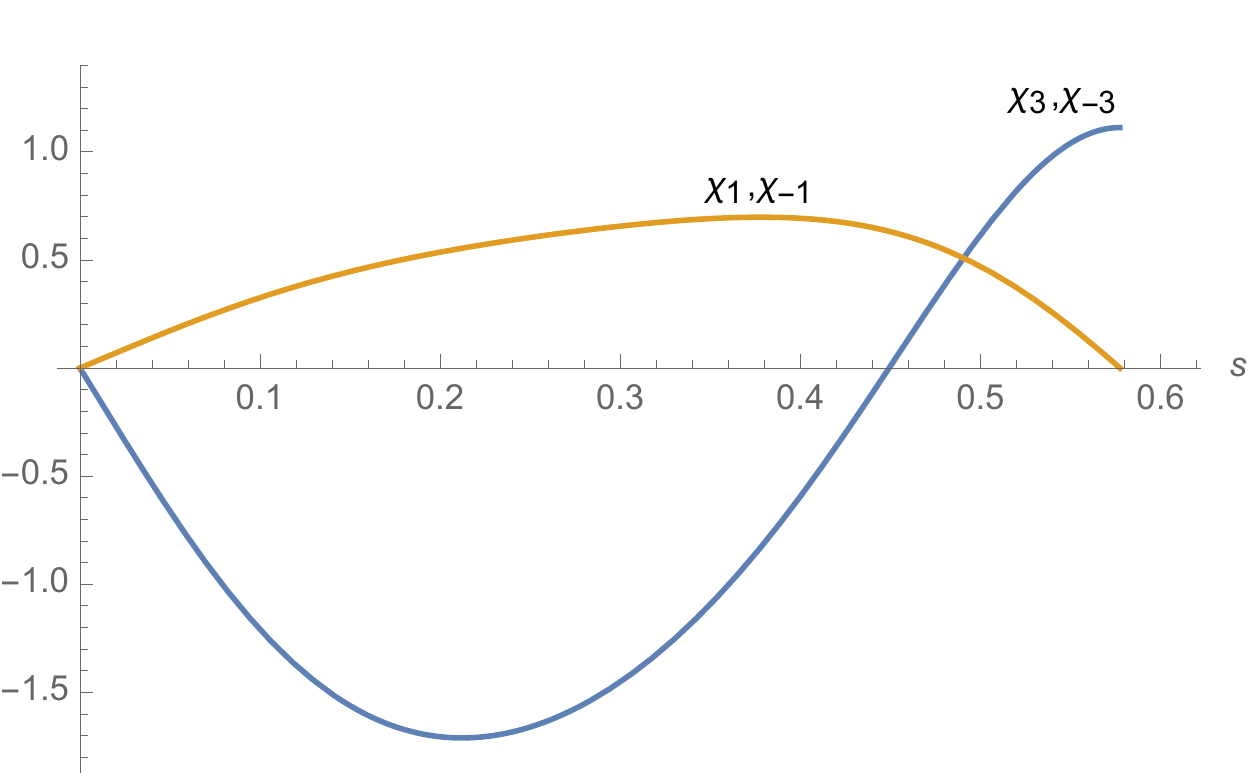} 
		
	\end{subfigure}
	\begin{subfigure}{0.49\textwidth}
		\includegraphics[width=\textwidth]{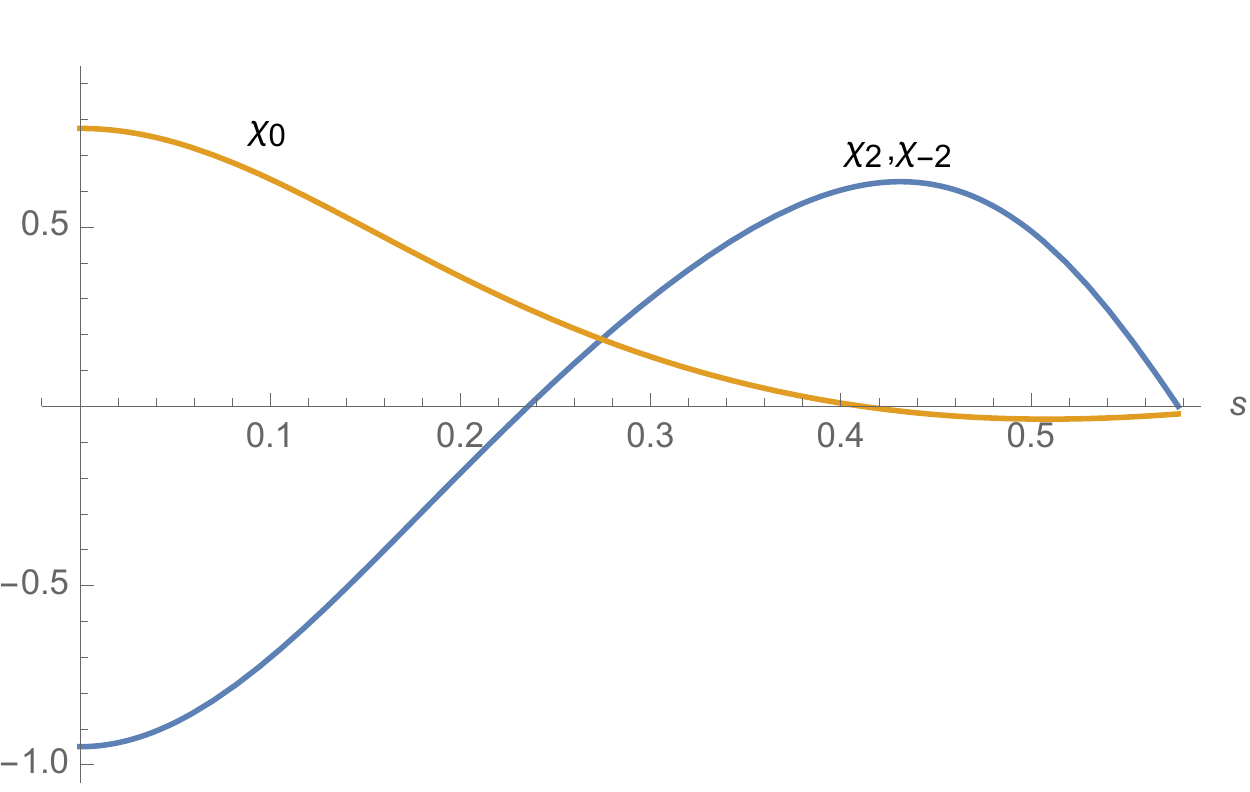} 
		
	\end{subfigure}

	\caption{Wavefunctions of the second \(3^-\) state and the third \(2^+\) state.}
	\label{fig9}
\end{figure}

This is a significant improvement on the rigid body spectrum previously discussed. Many observed states that were missing are now present. These include the low-lying \(1^-\) and \(2^-\) states as well as a second \(3^-\) state and a third \(2^+\) state slightly higher up (Figure \ref{fig9}). The spectrum also includes a \(3^+\) state and a \(4^-\), \(4^+\) pair (at energies 19.8 MeV, 21.3 MeV and 22.2 MeV), which have not yet been observed experimentally. There is an approximate correspondence between these states and the ACM rotational band labeled by \(\left(v_1,v_2^{l_2}\right) = \left(0,1^{1}\right)\), which has the same sequence of allowed spin and parity combinations. In the ACM picture, these states are associated with the doubly degenerate vibration of the equilateral triangle with \(E\) symmetry (with \(v_2 = 1\) denoting one unit of vibrational excitation).

\begin{figure}[h!]
	\centering
	
	\begin{subfigure}{0.44\textwidth}
		\includegraphics[width=\textwidth]{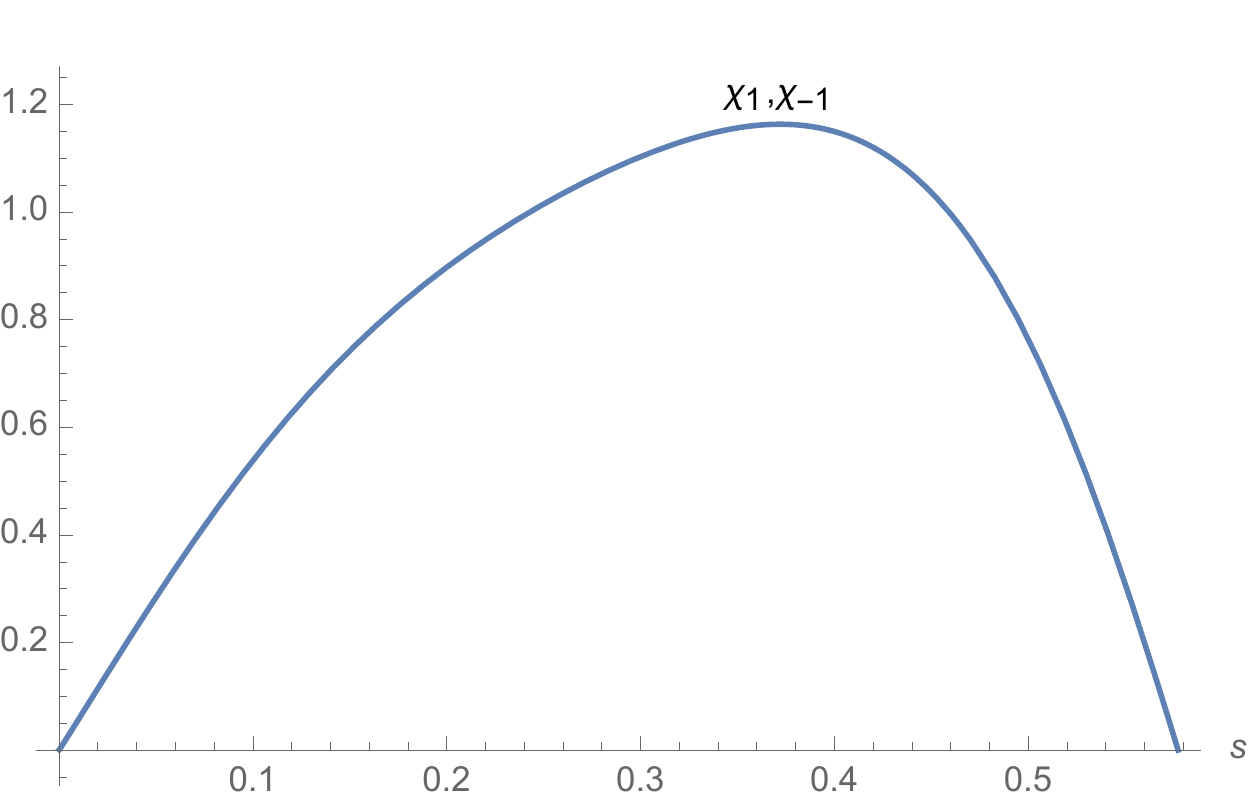} 
		\caption*{\(1^-\)}
	\end{subfigure}
	\begin{subfigure}{0.44\textwidth}
		\includegraphics[width=\textwidth]{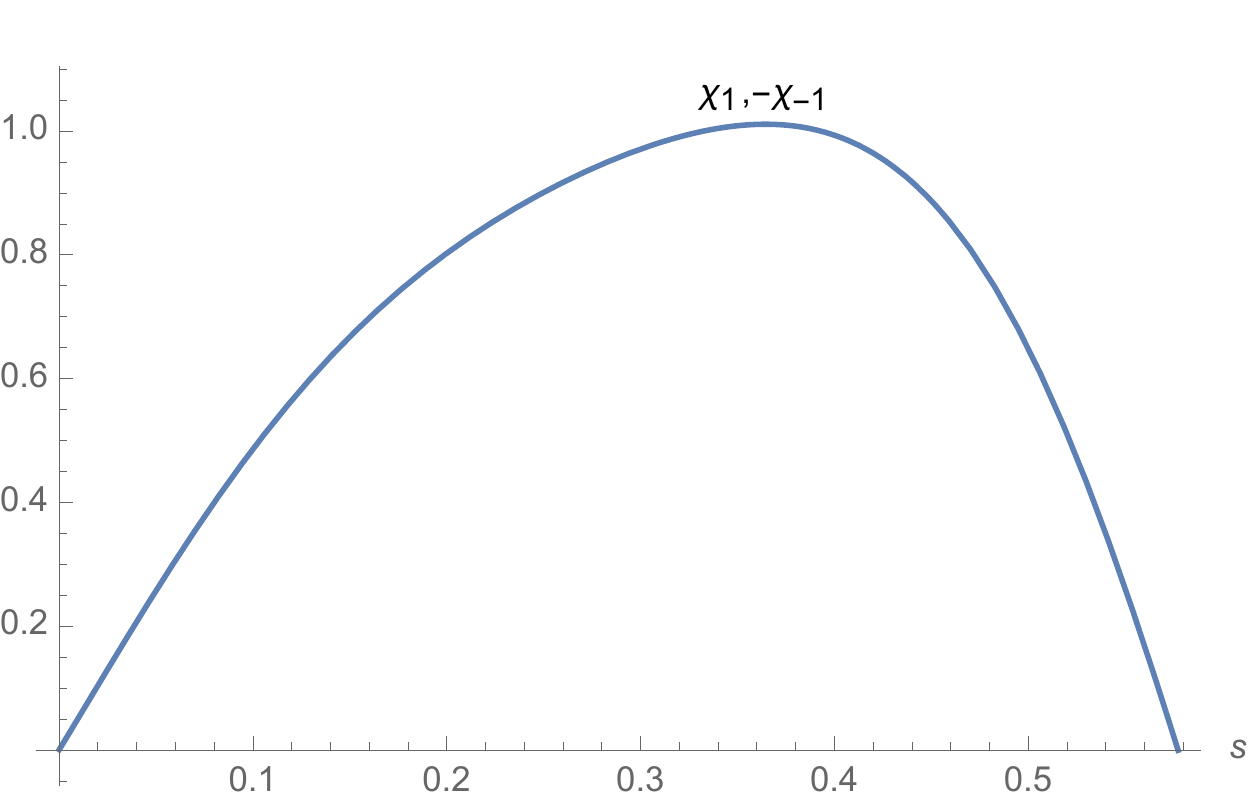} 
		\caption*{\(2^-\)}
	\end{subfigure}
	
	\begin{subfigure}{0.49\textwidth}
		\includegraphics[width=\textwidth]{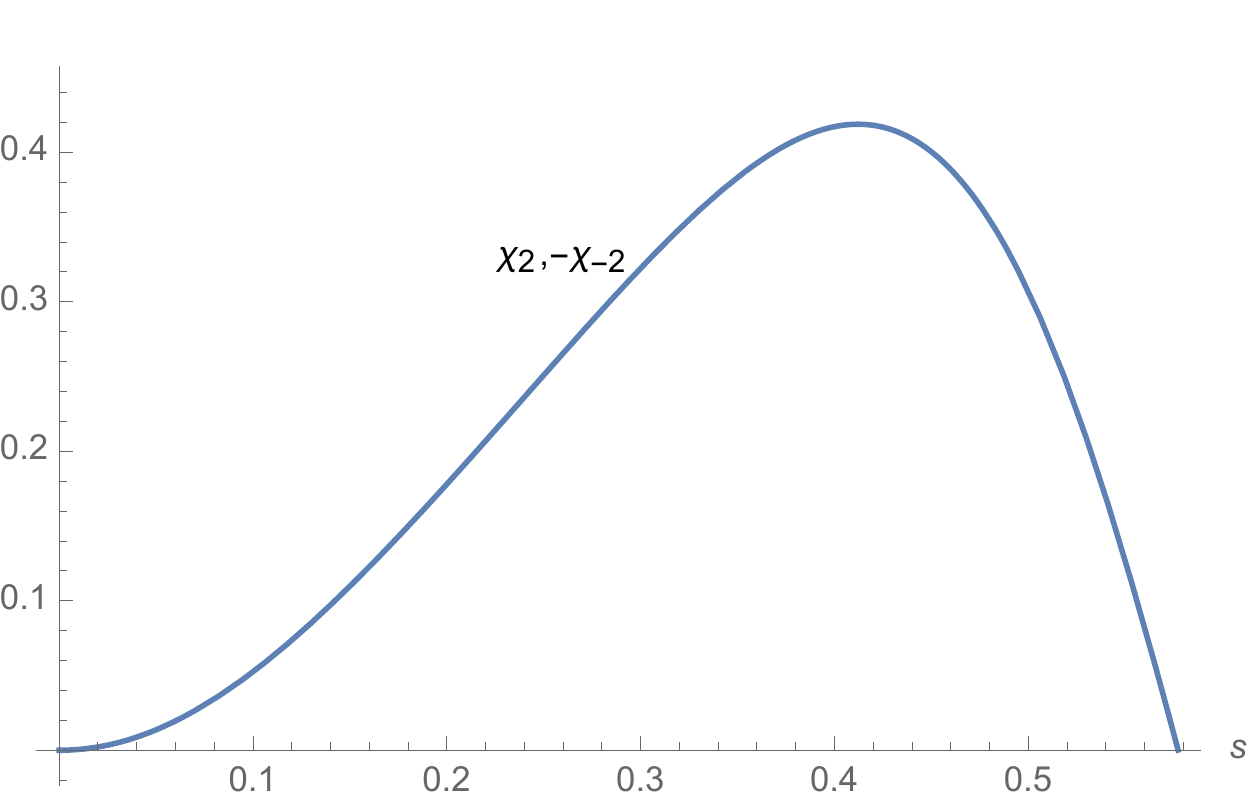} 
		\caption*{\(3^+\)}
	\end{subfigure}
	
	\caption{Wavefunctions of the lowest \(1^-\), \(2^-\) and \(3^+\) states.}
	\label{fig10}
\end{figure}

Our wavefunctions give further insight into the nature of the excited states. The  \(1^-\), \(2^-\) and \(3^+\) are concentrated at a bent arm configuration (Figure \ref{fig10}), a shape between the equilateral triangle and the linear chain in our configuration space. They vanish at the equilateral triangle and the linear chain. The different spins correspond to rotational excitations of such a state.

\begin{figure}[h!]
	\centering
	\begin{subfigure}{0.49\textwidth}
		\includegraphics[width=\textwidth]{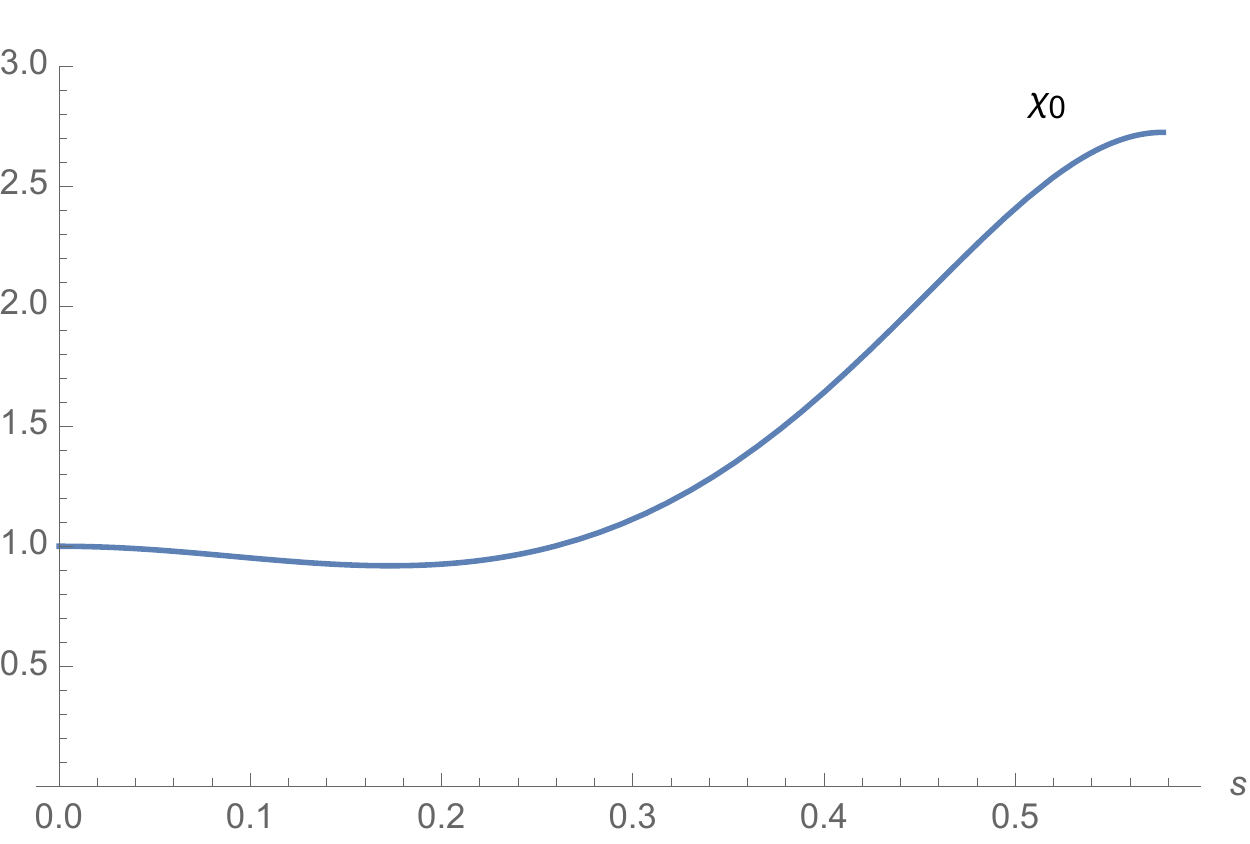} 
		
	\end{subfigure}
	\begin{subfigure}{0.49\textwidth}
		\includegraphics[width=\textwidth]{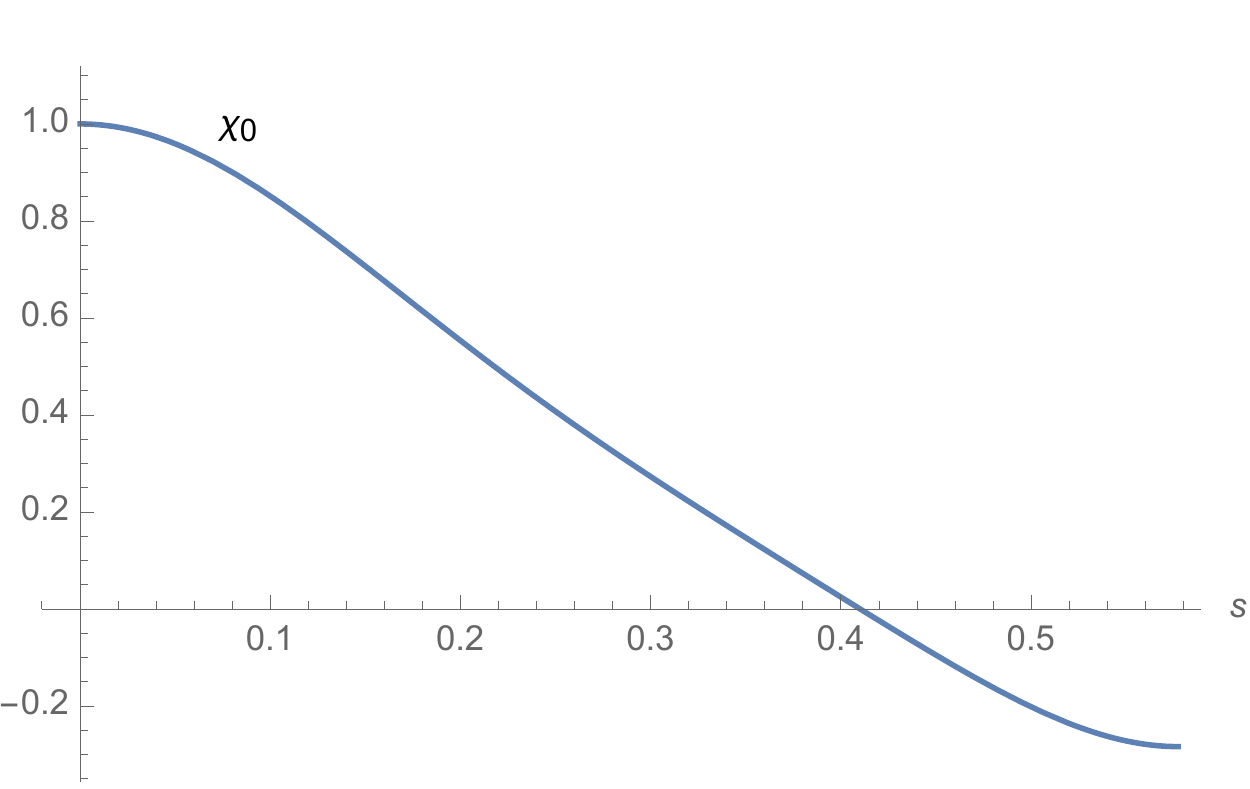} 
		
	\end{subfigure}

	\caption{Wavefunctions of the \(0^+\) ground state (left) and the \(0^+\) Hoyle state (right).}
	\label{fig11}
\end{figure}

The remaining low-lying states do involve the equilateral triangle and the linear chain. The wavefunctions of the \(0^+\) ground state and the \(0^+\) Hoyle state are plotted in Figure \ref{fig11}. It is interesting to compare these wavefunctions with the findings of recent lattice calculations \cite{epelbaum1} which suggest the \(0+\) ground state and the \(0^+\) Hoyle state have strong overlap with a compact triangular and a bent arm configuration respectively. Our wavefunctions are consistent with this picture: the ground state wavefunction peaks at the equilateral triangle and falls away fairly rapidly towards the linear chain, while the Hoyle state wavefunction peaks at the linear chain and is more spread out, remaining significant over a range of bent arm configurations.

\section{Conclusions}
We have improved upon a previous analysis which attempted to explain the low-lying spectrum of Carbon-12 in terms of a rigidly rotating equilateral triangle or linear chain \cite{mantonlau}. This analysis assumed a high degree of symmetry for the configurations relevant at low energies, leading to very few low-lying states and missing out several spin and parity combinations. By contrast, a model based on a local analysis of small vibrations leads to a spectrum with too many low-lying states \cite{bijkeriacello}. 

These problems are resolved once we allow shapes which interpolate between the equilateral triangle and the linear chain. The model presented in this paper, which utilises quantum graph theory, takes larger deformations seriously and gives us a different picture of the excited states. States which were considered independent in the rigid body picture can mix in our model, leading to superpositions of equilateral triangular and linear chain states. Other states which were not present in the rigid body picture have wavefunctions peaked at an intermediate bent arm configuration, rather than looking like a vibrational excitation of an equilateral triangle. We also predict three new energy levels for the Carbon-12 nucleus at around 20 MeV, with spin and parity combinations \(3^+\), \(4^-\) and \(4^+\).

The approach used in this paper is widely applicable, being of use in any situation where low-energy states have wavefunctions focused on particular lines in shape space of enhanced symmetry. For example, it would be worth revisiting previous work on quantisation of the \(B=7\) Skyrmion \cite{b7}. The quantum graph picture clarifies how one can consistently define a global wavefunction on a configuration space which is not a manifold (but a graph) and resolves difficulties which were encountered in that work.

\section*{Acknowledgements}

I am grateful to my supervisor Professor Nick Manton and also to Chris King and Chris Halcrow for many useful discussions and guidance. I would like to thank Dr David Foster for showing us numerical simulations of classically spinning \(B=12\) Skyrmions. I am supported by an EPSRC studentship.

\end{document}